\documentclass{jfm}
\usepackage{graphicx}
\usepackage{epstopdf, epsfig}
\usepackage{amsmath,amssymb, amsbsy, amsfonts, natbib}
\usepackage{bm,comment}
\usepackage{xcolor}
\usepackage{float}
\usepackage{caption}
\usepackage[hidelinks]{hyperref}

\newcommand{\addOBB}[1]{\textcolor{black}{#1}}

\newcommand{\cE}{\mathcal{E}}
\newcommand{\cA}{\mathcal{A}}
\newcommand{\cB}{\mathcal{B}}
\newcommand{\cP}{\mathcal{P}}

\newcommand{\quand}{\quad\mbox{and}\quad}
\newcommand{\qst}  {\quad\mbox{such that}\quad}

\newcommand{\qwr} {\quad\mbox{where}\quad}

\newcommand{\thf}{\quad\Rightarrow\quad}

\newcommand{\half}{\frac{1}{2}}
\newcommand{\bk}{\mathbf{k}}
\newcommand{\bu}{\mathbf{u}}
\newcommand{\bF}{\mathbf{F}}
\newcommand{\oma}{\omega}

\newcommand{\Oma}{\Omega}

\newcommand{\omi}{\hat\omega}

\newcommand{\Sta}{{St}_\alpha}

\newcommand{\dxdz}{\mathrm{d}x \mathrm{d}z}

\newcommand{\Za}{Z_\alpha}

\newcommand{\dert}[1]  {\frac{{ \mathrm{d}}{#1}}{{\mathrm{d}}{t}}}
\newcommand{\Dert}[1]  {\frac{{ \mathrm{D}}{#1}}{{\mathrm{D}}{t}}}

\renewcommand{\div}{\nabla\cdot}
\newcommand{\curl}{\nabla\times}

\newcommand{\unitv}[1] {\mathbf{\hat #1}}

\newcommand{\bka}{\mathbf{k}_\alpha}
\newcommand{\bkb}{\mathbf{k}_\beta}
\newcommand{\bkc}{\mathbf{k}_\gamma}

\newcommand{\bcg}{\mathbf{c_g}}
\newcommand{\grad}{\nabla}

\newcommand{\eps}{\varepsilon}

\newcommand{\ea}{e_\alpha}
\newcommand{\eb}{e_\beta}
\newcommand{\ec}{e_\gamma}

\renewcommand{\oma}{\omega_\alpha}
\newcommand{\omb}{\omega_\beta}
\newcommand{\omg}{\omega_\gamma}
\newcommand{\omc}{\omega_\gamma}

\shorttitle{Wave turbulence of inertia--gravity waves}
\shortauthor{Shavit, B\"uhler, Shatah}

\title{Wave turbulence of inertia--gravity waves: a theory for the oceanic spectrum}

\author{Michal Shavit\aff{1}\corresp{\email{ms14479@nyu.edu}}, Oliver B\"uhler\aff{1} 
and Jalal Shatah\aff{1}}

\affiliation{\aff{1}Courant Institute of Mathematical Sciences, New York University, NY 10012, USA
}

\begin{document}

\maketitle

\begin{abstract}
  We present a derivation using kinetic wave theory of the two-dimensional empirical Garrett--Munk spectrum for ocean internal waves, valid at all frequencies including near-inertial frequencies.  This is based directly on the governing equations for a two-dimensional Boussinesq system with constant stratification and rotation.  Our results improve on previous work by side-stepping the use of canonical variables, by taking full account of the Coriolis parameter in a non-hydrostatic dispersion relation, by filtering the balanced flow component from the dynamics, by using the conservation laws for energy and two components of pseudomomentum  to bring the collision integral  into a very simple form, by giving precise convergence conditions for the collision integral, and by finding  the unique scale-invariant turbulent wave spectrum that corresponds to turbulent fluxes from small to large wavenumbers. The last step was achieved in the limit of small but nonzero Coriolis parameter.      
  
  Key results are that any nonzero Coriolis parameter regularizes the singular nature of the non-rotating problem and that the homogeneity properties of the dispersion relation and of the  interaction coefficients alone already imply that the spectrum is separable in vertical wavenumber and frequency.   Within the restrictions of two-dimensional dynamics, this provides a theoretical framework for internal-wave turbulence consistent with oceanic observations.

\end{abstract}

\begin{keywords}
Internal waves, waves in rotating fluids, oceanography.
\end{keywords}

\section{Introduction}
\label{sec:intro}

We describe the importance of broadband spectra of
internal waves in the deep ocean and summarize the approach of
kinetic wave theory to model the formation of such spectra.  
This is followed by a discussion of the conceptual
obstacles that have made progress in this area difficult compared to other
applications of kinetic theory.  We then describe how our
recently developed approach has overcome most of these obstacles, albeit only
in  two spatial dimensions.   Detailed numerical results
for small but nonzero values of the Coriolis parameter
then produce two-dimensional wave spectra consistent with the
full frequency range of the observed spectra, including the
prominent peak at near-inertial frequencies.  

\subsection{Ocean internal waves}
\label{sec:ocean-internal-waves}

Internal gravity waves are a central channel of energy transfer
towards small-scale ocean mixing across density surfaces in the
deep ocean \citep{mackinnon2017climate, whalen2020internal},
which is well known to be crucial for the functioning of the
global ocean circulation on long time scales.  However, the
relevant waves are far too fast and far too small to be directly
resolvable by global circulation models, hence their dynamical
behaviour and their induced mixing must be parametrized using a
combination of theory and observations.  This necessitates a
statistical approach aimed at discerning average power spectra of
waves and in this connection much attention has been given to the
observation that broadband oceanic wave spectra are remarkably
similar in shape and amplitude across the ocean and resemble
quite closely the celebrated Garrett--Munk~(GM) spectrum
\citep{garrett1979internal, munk1981internal, ferrari2009ocean}.
For reference, this is a separable spectrum in $(\omega,m)$-space where $\omega\geq f$ is the frequency and $m$ is the vertical wavenumber, with typical form
\begin{equation}\label{eq:GMspec}
e_{GM}(\omega,m)\,\mathrm{d}\omega\mathrm{d}m = \frac{C_0}{ m^{2}} \frac{\mathrm{d}\omega\mathrm{d}m}{\omega\sqrt{\omega^2-f^2}}.
\end{equation}
Here $C_0>0$ is a constant and $f$ is the Coriolis parameter.  This form of the GM spectrum is valid at all frequencies but is limited to vertical wavelengths falling between $10-1000$ metres.  This is because at larger scales the finite depth of the ocean matters and at smaller scales the spectrum falls off rapidly due to dissipation by shear instability.       

The ubiquity of the GM spectrum arise despite the fact that the known principal sources of
internal waves due to wind forcing and tidal interactions with
rough topography are very narrow in frequency, peaking at the
inertial frequency and the frequency of the semi-diurnal tide,
respectively.  How such nearly monochromatic and spatially
inhomogenenous wave forcing mechanisms translate into a
reasonably universal broadband spectrum across the ocean has been
an enduring puzzle in physical oceanography ever since
\citep{olbers1983models}.  

\subsection{Kinetic wave theory}
\label{sec:kinetic-wave-theory}

One candidate to explain the emergence of broadband spectra from
narrowband forcing has been the kinetic theory of resonant
wave--wave interactions, which had been originally developed in
geophysics for surface waves and then adapted to internal waves
\citep[e.g.][]{hasselmann1966feynman,
  mccomas1977resonant,lvov2001hamiltonian,lvov2010oceanic,dema23,lanchon2023energy,labarre2024internal,labarre2024kinetics},
where resonant interactions occur on wave triads.  This kind of
wave turbulence theory assumes that the flow field consists of
many small-amplitude wave modes that interact weakly with each
other.  Most of these wave--wave interactions are non-resonant
and therefore have negligible impact on the wave spectrum, but a
subset of resonant interactions (in which a nonlinear combination
of wave modes can force another wave mode resonantly at its
precise frequency) can change the spectral amplitudes
significantly.  Remarkably, these assumptions can be turned into
a systematic turbulent closure scheme that leads to a nonlinear
integral equation called the kinetic equation, which describes
the slow nonlinear evolution of the energy spectrum in terms of a
collision integral that accounts for the resonant interactions within
the spectrum.  

The basic derivation and workings of this kind of
theory are by now well known and available in textbooks
\citep[e.g.][]{ZLF,nazabook,galtier2022physics}.  The most
important non-trivial outcome from these theories is the
prediction of unique power law exponents for the so-called
Zakharov--Kolmogorov spectra that describe wave energy cascades
across inertial wavenumber bands far away from forcing and
dissipation scales.  In contrast to the familiar Kolmogorov
spectra of standard, hydrodynamic turbulence, these power laws
can be derived from first principles based on the kinetic
equation.

\subsection{Obstacles specific to internal waves}
\label{sec:obst-progr-intern}

Despite this promising outlook it has been hard to make definite
predictions for the ocean internal wave case, for a number of
reasons.  First, the kinetic wave theory has been developed furthest
for systems that can be written in the variables
familiar from canonical Hamiltonian systems.  But the relevant
fluid systems are well-known to be non-canonical Hamiltonian
systems and do not fit into this description \citep{salmon1988hamiltonian}.  Second, theoretical
predictions are strongest for isotropic systems where no
direction is special for either the wave dispersion relation or
for the emerging wave spectrum; taking into account anisotropic
effects is reasonably straightforward only in the case of a
monomial dispersion relation such as $\omega=k^am^b$, where
$\omega$ is the wave frequency, $(k,m)$ are components of a
multidimensional wavenumber vector, and $(a,b)$ are some
exponents.    But this is not true for the famous internal wave dispersion relation
(see \S\ref{sec:linlin})
\begin{equation}\label{eq:omegafirst}
\omega^2 = \frac{N^2k^2+f^2m^2}{k^2+m^2},
\end{equation}
where $N$ is the buoyancy frequency, $f$ is the Coriolis
parameter, and $(k,m)$ are horizontal and vertical wavenumbers.
This has prompted attempts to approximate it with a monomial form
in certain limits, for example by ignoring $f^2$ and making the
hydrostatic approximation $\omega^2\approx N^2k^2/m^2$.  Notably,
whilst this is accurate for some parts of the ocean wave
spectrum, it is also wildly inaccurate for other parts, and the
methods of kinetic wave theory are global in nature and
necessarily involve the entire spectral range. At zero rotation, $k=0$ corresponds to slow modes, their inclusion is incompatible with fundamental assumption in the derivation of the kinetic equation, like Gaussianity of the initial distribution of waves.
Third, the oceanic internal wave field is strongly affected by Coriolis
forces, i.e., the relevant waves are inertia--gravity waves,
which feel both the stratification $N$ and the rotation $f$.  The
presence of two distinct important dynamical mechanism is an
additional complication, which for example thwarts attempts at a
monomial approximation as described before.  Fourth, with
rotation present it also becomes harder to describe flows that
consists only of internal waves and contain no balanced vortex
motion.  This is a condition that is most fundamentally linked to
the absence of gradients in the potential vorticity~(PV) of the
flow, a nontrivial criterion if $f\neq0$.

\subsection{Results from our work}
\label{sec:results-from-our}

In our work on this problem, which is reported in
\citep{shavit2024sign,shavit2025turbulent} and the present paper,
we have made fundamental advances on all of the above four
points, and indeed we think we have  solved the problem for
two-dimensional turbulent inertia--gravity waves in the regime $f\ll N$.  This
progress comes at the price of restricting our study to two-dimensional
motion in a vertical plane, which has a number of conceptual and
numerical advantages.  Arguably, there is no obvious three-dimensional
effect that is missed by doing so, hence we believe this is a
useful step towards solving the problem more generally.
Nonetheless, this restriction must be clearly noted.

On the first point we follow \cite{ripa} and sidestep any
consideration of canonical variables by basing our theory
directly on an expansion of the governing equations in terms of
the normal modes of the linear system, which form a complete set
for the full sytem and are orthogonal with respect to the total
energy.  This allows a straightforward formulation of the exact
evolution of the modal amplitudes and makes any quadratic
conservation laws completely explicit.  We then exploit the
inherent conservation laws for energy and pseudomomentum in our
system to obtain a kinetic equation that is of a comparatively
simple form and amenable to further analysis.  On the second
point we allow for the full dispersion relation and for arbitrary
angular dependence of the spectrum.  Following the important
methods developed in \cite{balk2000kolmogorov}, this is achieved
by exploiting only that the dispersion relation and the kinetic
equation are homogeneous in their dependence on the wavenumbers.
On the third point we are able to take full account of the
Coriolis force by considering exact conservation laws for both the
horizontal and vertical components of pseudomomentum, which is
novel in this area.  Finally, for the fourth point we explicitly
restrict to an approximate wave manifold by setting the linear PV to zero
at the outset.  We show that this has some repercussions for the general conservation
laws but does not affect the resonant interactions relevant for
the kinetic equation.    

Based on this approach we showed in \cite{shavit2024sign} that
the only power law candidate for a turbulent inertial-range spectrum in the non-rotating
two-dimensional system is a spectral density of the form
$e(k,m)=C_0K^{-3}g(\theta)$, which uses polar coordinates
$(k,m)=K(\cos\theta,\sin\theta)$.  This was achieved by a
detailed analysis of the collision integral, which was assumed to
converge for these kinds of spectra.   Notably, it was not
necessary to find the correct angular dependence $g(\theta)$ at
this stage.  This paper also described the formal extension of
the kinetic equation to the
rotating case, which is however much more complex than in the
non-rotating case.   The follow-up study in
\cite{shavit2025turbulent} was aimed at finding the relevant
angular dependence $g(\theta)$ in the non-rotating case. The presence of zero-frequency modes when $f=0$ and
$k=0$ in \eqref{eq:omegafirst} requires considering a regularized frequency domain for the kinetic equation to hold. This was achieved by cutting out a small sector of near-zero frequencies; we then searched for the turbulent spectrum on this regularized domain.
%\addMS{However, the presence of zero-frequency modes when $f=0$ and
%$k=0$ in \eqref{eq:omegafirst} led to severe divergences} in the
%kinetic equation and progress required cutting out a small sector
%of near-zero frequencies and searching the collision integral
%numerically for turbulent spectra in this regularized domain.
The width of the frequency cut-out was then reduced iteratively
and very clear numerical convergence to a turbulent spectrum of
the form
\begin{equation}
  \label{eq:21}
  e(k,m)=C_0K^{-3}\omega(\theta)^{-2}
\end{equation}
was found.  This was a new
result, different from all previous two-dimensional spectra
computed before.  Remarkably, rewritten as a density over
$(m,\omega)$-space this spectrum is exactly proportional to
$m^{-2}\omega^{-2}$, which is the form of the GM spectrum (\ref{eq:GMspec}) for large vertical
wavenumber and frequency. To our knowledge, this was the first
time that a two-dimensional spectrum compatible with the
two-dimensional GM spectrum had been derived from kinetic
wave theory. A subsequent DNS study of the 2D Boussinesq equation without rotation confirmed this spectrum in the weak wave turbulence regime \citep{labarre20242d}.

In the present study we allow for $f\neq0$ and present a much
more detailed study of the rotating collision integral.
Crucially, $|\omega|$ is now bounded between $N$
and $f$ and hence strictly nonzero, which eliminates
zero-frequency waves.  In other words, the presence of rotation,
no matter how weak, provides a physical frequency cut-off that
eliminates the divergences of the non-rotating case.  This now
allows determining precise convergence criteria for the collision
integral, which previously had to be assumed to converge just
formally.  The price to pay for this is the significantly more
complex nature of the collision integral in the rotating case.
We were not able to discern a turbulent inertial-range spectrum
for arbitrary values of $f/N$, but for very small values very
good numerical evidence was found that the spectrum was again
given by \eqref{eq:21}.
Transforming to  a spectral density in $(m,\omega)$-space yields
\begin{equation}
  \label{eq:22}
  C_0K^{-3}\omega(\theta)^{-2}\,\mathrm{d}k\mathrm{d}m \propto
  m^{-2}\,\frac{\mathrm{d}\omega\mathrm{d}m}{\omega\sqrt{\omega^2-f^2}},
\end{equation}
which is the GM spectrum with a full frequency dependence valid
for all frequencies, including near-inertial frequencies.  To our
knowledge this is the first time that this has been achieved.
We also show that separability of the turbulent spectrum density 
in $m$ and $\omega$ is already implied by the homogeneity
with respect to wavenumbers of the dispersion relation and the
kinetic equation.  This underpins another key property of the
empirical GM spectrum.

The plan of the paper is as follows.  The governing equations, 
conservation laws, modal expansions for the waves and their resonant interactions are 
described in \S\ref{sec:gove}.   The kinetic equation and its
formal properties are formulated in \S\ref{sec:kestuff}, which
is then followed by a detailed study of the convergence of the
collision integral.  The numerical search for a turbulent
spectrum is then described in \S\ref{sec:turbulent-spectrum} and
\S\ref{sec:angpart}.  Some concluding comments are offered in
\S\ref{sec:concl} and details about the parametrization of the
resonant manifold are offered in an appendix.

\section{Governing equations} 
\label{sec:gove}

The two-dimensional Boussinesq system in a vertical $xz$-slice
derives from the full Boussinesq system by setting all
$y$-derivatives to zero. In the presence of rotation the velocity
field then remains three-dimensional, i.e., there is a nonzero $v$-velocity in/out of
the $xz$-plane.  The governing equations can be written in a
non-standard but useful fashion as 
\begin{equation}\label{eq1a}
\Dert{u}+P_x =fv,\quad \Dert{w}+P_z = b,\quad u_x+w_z=0
\end{equation}
and 
\begin{equation}\label{eq1b}
\Dert{(fv)}+f^2u =0,\quad \Dert{b}+N^2w = 0.
\end{equation}
Here $\bu=(u,v,w)$ are the velocity components in the
$(x,y,z)$-directions, $b$ is the buoyancy, $P$ is the pressure
divided by a reference density, $f$ is the Coriolis parameter,
and $N$ is the buoyancy frequency.  Both $f$ and $N$ are assumed
constant and  $N>f>0$.  The typical oceanic regime is
$N/f\geq 40$.

Equation \eqref{eq1b} stresses the similarity between
the buoyancy force $b$ in the vertical and the inertial force $fv$ in the horizontal.  Indeed,
the two-dimensional equations are symmetric under the ninety degree rotation
\begin{equation}\label{eq:thesym}
(x,z)\to(z,-x),\quad (u,w)\to(w,-u),\quad (fv,b)\to (b,-fv),\quad (f,N)\to(N,f).
\end{equation}
This exact symmetry does not hold for the full, three-dimensional
Boussinesq system \citep{veronis1970analogy}.  Consequently,
there are two materially invariant fields, namely
\begin{equation}
\label{eq2}
S=N^2z+b\quand M=f^2x + fv \qst\Dert{S} = \Dert{M} = 0.
\end{equation}
Hence the level curves of $S$ and $M$ form a curvilinear mesh
that is materially advected, which automatically visualizes the flow map in the $xz$-plane.      Moreover, the exact potential vorticity~(PV) is
\begin{equation}
\label{eq3}
Q = (\curl\bu + f\unitv{z})\cdot\nabla(N^2z+b) = fN^2 + fb_z +N^2 v_x + v_xb_z-v_zb_x = \frac{1}{f}J(M,S)
\end{equation}
where the Jacobian $J(M,S)=M_xS_z-M_zS_x$.  Hence the PV is
proportional to the area element of the materially advected
curvilinear mesh.  If the PV is constant then the $(M,S)$-mesh
has uniform area content; this is the case relevant for
flows consisting entirely of internal waves.  Below we use
\begin{equation}
q=(Q-fN^2)/N^2 = v_x + \frac{f}{N^2}b_z + \frac{1}{N^2}J(v,b),
\end{equation}
which has units of vorticity and is zero if the PV is equal to its background value $fN^2$.  
Overall, we view our system as describing the
dynamics of an area-preserving two-dimensional velocity field
$(u,w)$ subject to a two-dimensional restoring body force $(fv,b)$ due to
rotation and stratification.

\subsection{Conservation laws}

With suitable boundary conditions the system conserves the familiar quadratic energy 
\begin{equation}
\label{eqE}
\cE = \frac{1}{2} \int \left(u^2+v^2+w^2+\frac{b^2}{N^2}\right)\,\dxdz.
\end{equation}
Notably, $v^2=(fv)^2/f^2$ could also be viewed as part of the
available potential energy on equal footing with $b^2/N^2$.  It is  well known
\citep{scinocca1992nonlinear} that the non-rotating system 
conserves the horizontal pseudomomentum
\begin{equation}
\label{eqPx}
\cP_x = \int \frac{b(u_z-w_x)}{N^2}\,\dxdz.
\end{equation}
Less familiar is that $\cP_x$ is conserved also in the presence
of rotation, but then only if the PV is uniform, i.e., if $q=0$
everywhere \citep{buhler1999shear}.  Indeed, direct manipulation
leads to
\begin{equation}\label{eq:pxc}
\Dert{}\left(\frac{b(u_z-w_x)}{N^2}\right) + \div\bF + vq = 0
\end{equation}
where the flux vector
\begin{equation}
\bF = \left(\half\left(\frac{b^2}{N^2}+u^2-v^2-w^2-\frac{v^2b_z}{N^2}\right), \, uw-\frac{f}{N^2}bv+\frac{v^2b_x}{2N^2}\right).
\end{equation}
It follows from the symmetry (\ref{eq:thesym}) that a vertical pseudomomentum in the form
\begin{equation}
\label{eqPz}
\cP_z = -\int \frac{v(u_z-w_x)}{f}\,\dxdz
\end{equation}
is also exactly conserved under the same condition, which does
not seem to have been recognized before.  Specifically,
conservation demands that the source term $vq$ in \eqref{eq:pxc}
integrates to zero.  Integration by parts yields
\begin{equation}\label{eq:source}
vq = \frac{1}{2} (v^2)_x + \frac{f}{N^2}vb_z +
\frac{1}{2N^2}J(v^2,b) = \ldots -\frac{f}{N^2}bv_z, 
\end{equation}
where the final term is the only term that does not integrate to
zero automatically in a periodic domain.   This term  is linked intrinsically to rotation.

\subsection{Linear modes}
\label{sec:linlin}

In a periodic domain linear modes are found by demanding that all
fields are proportional to $\exp(i[kx+mz-\omega t])$ and
substituting into the linearized equations; these modes form a
complete set in which to decompose the flow at any given moment.
The dispersion relation has three roots.  One is a zero root
$\omega=0$ describing a steady flow.  This mode has no
in-plane velocity $u^{b}=w^{b}=0$ but $fv^{b}=P^{b}_x$ and
$b^{b}=P^{b}_z$, i.e., there is geostrophic and hydrostatic
balance.  This is why this steady mode is commonly referred to
as the balanced mode.  The linear PV
$q^b$ for the balanced mode is therefore 
\begin{equation}
  q^b = v^{b}_x +\frac{f}{N^2}b^{b}_z = \frac{1}{f}\left(P^b_{xx}
    + \frac{f^2}{N^2}P^b_{zz}\right).
\end{equation}
In a periodic domain this establishes a uniquely invertible map
between  zero-mean functions $q^b$ and $P^b$, hence the
balanced mode can be diagnosed from the linear PV.  In
particular, the balanced mode is zero if $q^b=0$ throughout.  The
balanced mode has nonzero energy but zero pseudomomentum.
In two-dimensional flow the linear balanced mode is also a nonlinear
steady solution, but this is not true in three dimensions.

Linear internal waves obey the homogeneous dispersion relation
\begin{equation}\label{eq:omega}
\omega^2 = \frac{N^2k^2+f^2m^2}{k^2+m^2} = N^2\cos^2\theta +f^2\sin^2\theta,
\end{equation}
where the second form uses polar coordinates
$\bk=K(\cos\theta,\sin\theta)$.  It will be crucially important that
\begin{equation}
  \label{eq:8}
  N^2\geq \omega^2\geq f^2 > 0,
\end{equation}
i.e., that $|\omega|$ is bounded and strictly
positive.    We explicitly  account for the equal-and-opposite
branches of (\ref{eq:omega}) by writing 
\begin{equation}\label{eq:disp}
\omega = \sigma\,\hat\omega \qwr \hat\omega = + \sqrt{N^2\cos^2\theta +f^2\sin^2\theta}
\end{equation}
and $\sigma=\pm1$ accounts for the sign of
$\omega$.\footnote{This convention differs from
  \cite{shavit2024sign,shavit2025turbulent}, where for $f=0$ the
  branch choice could be linked to left- or right-going waves in
  a continuous fashion.  This is not possible if $f\neq0$.}
Internal waves have nonzero pseudomomentum in general and a plane
internal wave integrated over a wavelength yields the famous
relation
\begin{equation}\label{eqpmE}
(\cP_x,\cP_z) = \frac{(k,m)}{\omega}\,\cE
\end{equation}
between the pseudomomentum vector and the wave energy.  The
linear PV of a wave mode 
\begin{equation}\label{eq:nowpv}
q^w = v^{w}_x +\frac{f}{N^2}b^{w}_z = 0
\end{equation}
for all $\bk=(k,m)$.

An important  consequence of (\ref{eq:nowpv}) is that the total
energy is orthogonal across all wave modes \textsl{and} across all
balanced modes.  This is because after integration by parts the mixed energy terms
\begin{equation}
v^{b}v^{w}+\frac{1}{N^{2}}b^{b}b^{w} = P^b_{x}v^{w}/f+\frac{1}{N^{2}}P^b_{z}b^{w} = \ldots 
 - P^b/f\left(v^{w}_x +\frac{f}{N^2}b^{w}_z\right) 
\end{equation}
integrate to zero due to \eqref{eq:nowpv}.   Hence at any moment
the total energy can be uniquely partitioned into wave and balanced
components, at least according to linear theory.  Unfortunately, this is not true for
the total pseudomomentum vector, where mixed
terms such as
\begin{equation}
\frac{b^{b}}{N^{2}}(u^{w}_{z}-w^{w}_{x}) = \frac{1}{N^{2}}P^b_{z}(u^{w}_{z}-w^{w}_{x})
\end{equation}
need not vanish upon integration.  Physically, this lack of
orthogonality when waves and balanced modes overlap in physical
space is linked to the lack of pseudomomentum conservation in the
presence of a balanced mode.  This can be demonstrated by
a thought experiment: consider the linear dynamics of a compactly supported
wavepacket to which a localized balanced flow has been added such
that at $t=0$ the buoyancy $b=b^{b}+b^{w}=0$.  This means that the horizontal
pseudomomentum $\cP_{x}=0$ initially.  Yet, as the wavepacket
propagates away it will manifest a nonzero $\cP_{x}$ given by
(\ref{eqpmE}), whilst the steady balanced mode is left behind and
has zero pseudomomentum.  So pseudomomentum is neither
conserved nor additive over wave and balanced modes if
the wavepacket overlaps with the balanced mode.

\subsection{Restriction to wave manifold}

We are interested in the long-term dynamics of weak internal
waves in isolation, i.e., in the absence of an additional
balanced flow.  Strictly speaking, we want to enforce the
nonlinear PV constraint that $q\equiv0$ throughout the domain,
which would then be conserved by the nonlinear dynamics.  For
small-amplitude, weak internal waves the best we can do is to
eliminate the balanced flow component according to linear theory
\textsl{a priori} by setting $q^b=0$.  Hence we only allow wave
amplitudes to be nonzero and we seek to track only  
wave--wave interactions, which is our definition of a 
restriction to a wave manifold.

We use greek indices $\alpha=(\sigma_\alpha,\bk_\alpha)$ to
identify wave modes and then the reality condition implies that
$\alpha$ and $-\alpha=(-\sigma_\alpha,-\bk_\alpha)$ refer to the
same physical wave mode.  Following \cite{ripa},
the complex wave amplitudes $Z_\alpha(t)$ for the full, nonlinear
system evolve as
\begin{equation}\label{eq:Zdot}
\dot Z_\alpha + i\oma Z_\alpha= \frac{1}{2}\sum_{\beta,\gamma}V_\alpha^{\beta\gamma}Z^*_\beta Z^*_\gamma.
\end{equation}
Here $V_\alpha^{\beta\gamma}=V_\alpha^{\gamma\beta}$ are real-valued wave--wave interaction coefficients that are nonzero only if the triad conditions
\begin{equation}\label{eq:kres}
k_\alpha+k_\beta+k_\gamma=0 \quand m_\alpha+m_\beta+m_\gamma=0
\end{equation}
hold for the 2d wavenumber vectors $\bk=(k,m)$.  Hence the
wavenumber vectors of interacting triads form triangles.  The
amplitudes satisfy the reality condition
$Z_{-\alpha}=Z^{*}_{\alpha}$ and are scaled such that
$E_\alpha=|\Za|^2=E_{-\alpha}$ is the energy in wave mode
$\alpha$.  By a standard argument of Galerkin truncation the quadratic energy is
conserved for each triad separately
\citep[e.g.][]{salmon1998lectures}.  This implies
\begin{equation}\label{eq:V}
V_\alpha^{\beta\gamma} + V_\beta^{\alpha\gamma} + V_\gamma^{\alpha\beta} =0.
\end{equation}
For simplicity, we write $V_\alpha=V_\alpha^{\beta\gamma}$ from
here on when it is clear that we are considering a specific triad
$(\alpha,\beta,\gamma)$.   By using the dispersion relation we
can rewrite (31) from \cite{shavit2024sign} in the beautifully
symmetric exact form
 \begin{equation}\label{Vdef}
V_\alpha = \frac{\mathbf{k_\beta}\times\mathbf{k_\gamma}}{K_\alpha K_\beta K_\gamma}\,\left\{\frac{N^2}{2\sqrt{8}}\left(s_\alpha + s_\beta + s_\gamma\right)\left(s_\beta-s_\gamma\right)+\frac{f^2}{2\sqrt{8}}\left(r_\alpha + r_\beta + r_\gamma\right)\left(r_\beta-r_\gamma\right)\right\}.
\end{equation}
Here $K=|\bk|$,
$\mathbf{k_\beta}\times\mathbf{k_\gamma}=\mathbf{k_\gamma}\times\mathbf{k_\alpha}=\mathbf{k_\alpha}\times\mathbf{k_\beta}$
is twice 
the signed area content of the triangle
\eqref{eq:kres}, and $(s,r)=(k,m)/\omega$ are the components of
the `slowness' vector.  This expression is still valid for all triads,
i.e., there has been no use of a frequency resonance condition so
far.  Only the slowness differences are affected by cyclic
permutations of $(\alpha,\beta,\gamma)$, which immediately proves \eqref{eq:V}.

Now, in the non-rotating case of \cite{shavit2024sign} the horizontal
pseudomomentum $s E$ was also conserved for every wave triad
separately, which led to
\begin{equation}\label{eq:sV}
\quad s_\alpha V_\alpha + s_\beta V_\beta + s_\gamma V_\gamma = 0.
\end{equation}
This holds from \eqref{Vdef} if $f=0$ but not otherwise.  
Similarly, conservation of vertical pseudomomentum $rE$ implies 
\begin{equation}\label{eq:rV}
r_\alpha V_\alpha + r_\beta V_\beta + r_\gamma V_\gamma = 0,
\end{equation}
which holds only if $N=0$ and not otherwise.  What this
manifests is that, unlike energy conservation, exact
pseudomomentum conservation would require carrying modal
amplitudes for the balanced modes as well, because of the lack of
orthogonality discussed before.  However, this unsatisfactory
situation is fully resolved when we restrict to resonant triads.

 \subsection{Resonant triads}
 \label{sec:resonant-triads}
 
We now restrict to resonant wave triads 
\begin{equation}
  \label{eq:1}
  \oma+\omb+\omega_\gamma = 0
\end{equation}
and show that for these triads the three constraints 
(\ref{eq:V}, \ref{eq:sV}, \ref{eq:rV}) are all satisfied
simultaneously.  Indeed, this overdetermined system has a
non-trivial solution of the form 
\begin{equation}
   \label{eq:2}
   \frac{V_\alpha}{\oma} =    \frac{V_\beta}{\omb} =    \frac{V_\gamma}{\omega_\gamma}, 
 \end{equation}
 as then (\ref{eq:sV}) and (\ref{eq:rV}) reduce to the triangle
 conditions (\ref{eq:kres}).  To demonstrate \eqref{eq:2} from
 \eqref{Vdef} we consider only the $N^2$ terms, as the $f^2$
 terms follow the same scheme.  Consider
\begin{equation}
  \label{eq:3}
  \frac{V_\alpha}{\oma}  = \frac{\mathbf{k_\beta}\times\mathbf{k_\gamma}}{K_\alpha K_\beta K_\gamma}\,\left\{\frac{N^2}{2\sqrt{8}}\left(s_\alpha + s_\beta + s_\gamma\right)\frac{\left(s_\beta-s_\gamma\right)}{\oma}\right\}
\end{equation}
and then
\begin{equation}
  \label{eq:4}
  \frac{s_\beta-s_\gamma}{\oma} =
  \frac{\frac{k_\beta}{\omb}-\frac{k_\gamma}{\omega_\gamma}}{\oma} =
  \frac{k_\beta\omega_\gamma- k_\gamma\omb}{\oma\omb\omega_\gamma}.
\end{equation}
This is the first component
of the three-dimensional cross product
\begin{equation}
  \label{eq:5}
  (k_\alpha,k_\beta,k_\gamma)\times(\oma,\omb,\omega_\gamma)
\end{equation}
times a factor that is totally symmetric under
permutations of $(\alpha,\beta,\gamma)$.
Repeating this for the other two components yields
\begin{equation}
  \label{eq:6}
  \left(\frac{V_\alpha}{\oma} ,   \frac{V_\beta}{\omb},
    \frac{V_\gamma}{\omega_\gamma}\right) = A \,(k_\alpha,k_\beta,k_\gamma)\times(\oma,\omb,\omega_\gamma)
\end{equation}
with some totally symmetric $A$.  
Now, for resonant triads both $(k_\alpha,k_\beta,k_\gamma)$ and $(\oma,\omb,\omega_\gamma)$ are
orthogonal to $(1,1,1)$ and therefore their
three-dimensional cross product is parallel to $(1,1,1)$, which
proves (\ref{eq:2}).  The $f^2$ terms follow from the same argument,
with vertical wavenumbers $m$ replacing horizontal
wavenumbers $k$.  The upshot is that for resonant triads 
\begin{equation}
  \label{eq:7}
  (V_\alpha,V_\beta,V_\gamma) = (\oma,\omb,\omega_\gamma)\,\Gamma
\end{equation}
holds with a 
\begin{equation}
  \label{eq:16}
  \Gamma =
  \frac{\mathbf{k_\beta}\times\mathbf{k_\gamma}}{K_\alpha K_\beta
    K_\gamma}\,\left\{\frac{N^2}{2\sqrt{8}}\left(s_\alpha +
      s_\beta +
      s_\gamma\right)\frac{\left(s_\beta-s_\gamma\right)}{\oma}+
    \frac{f^2}{2\sqrt{8}}\left(r_\alpha + r_\beta +
      r_\gamma\right)\frac{\left(r_\beta-r_\gamma\right)}{\oma}\right\} 
\end{equation}
that is totally symmetric in $(\alpha,\beta,\gamma)$, despite
appearance to the contrary.

\section{The kinetic equation and the turbulent spectrum}
\label{sec:kestuff}

In \cite{shavit2024sign}, we  derived a kinetic equation that
describes the slow evolution of the averaged wave energy density
$e_\alpha(t)=e(\sigma_\alpha,\bk_\alpha,t)=\left\langle E_{\alpha}\right\rangle$, assuming
$0<f<N/2$ and excluding the balanced modes. The brackets denote
averaging over an initial Gaussian statistical ensemble,
$\left\langle Z_\alpha^*(0) Z_\beta(0)\right\rangle =
\delta_{\alpha\beta} e_\alpha(0)$. The joint kinetic limits of
large domain and long nonlinear times were taken, yielding the
kinetic equation
\begin{align}
      \label{eq:kin sim}
  \partial_t \ea = St_\alpha ={\pi}\!\!
\int\limits_{\omega_{\alpha\beta\gamma}}\!\!\oma\,\Gamma^2 (\oma\eb\ec + \omb\ea\ec +
    \omg\ea\eb).%\delta(\omega_{\alpha \beta\gamma })
\end{align}
In this limit the discrete sum in \eqref{eq:Zdot} is replaced by an integral over the resonant manifold:
\begin{equation}\label{eq:resonant manifold}
    \int\limits_{\omega_{\alpha\beta\gamma}} \!\!\!\!\!= \!\int \mathrm{d}\beta \mathrm{d}\gamma \,\,\delta(\omega_\alpha+\omega_\beta+\omega_\gamma)\delta\left(\mathbf{k}_{\alpha}+\mathbf{k}_{\beta}+\mathbf{k}_{\gamma}\right),
\end{equation}
where
$\int\!\! \mathrm{d}\beta=\sum_{\sigma_\beta=\pm 1}
\int\!\!\mathrm{d}\mathbf{k_\beta}$, etc.  The collision integral
is over all modes $(\beta,\gamma)$ that can form resonant triads
with $\alpha$.  The reality
condition implies $e_\alpha=e_{-\alpha}$ so we consider only $\sigma_\alpha=1$ and therefore
$\oma>0$ throughout.  For $\beta$ and $\gamma$ both branches are relevant,
but we can use $e(-1,\bk,t)=e(+1,-\bk,t)$ as needed to express
everything in terms of  $e(\bk,t) \equiv e(+1,\bk,t)$.  In summary, the energy
spectrum is described by a single function $e(\bk,t)\geq0$
associated with the positive frequency branch $\omega>0$ and
shorthands like $\eb$ refer to $e(\sigma_\beta\bk_\beta,t)$.

\subsection{Formal properties of the kinetic equation}
\label{sec:form-prop-coll}

We enumerate a number of useful standard facts about (\ref{eq:kin sim})
that follow easily if the convergence of $St_\alpha$ and of $\int
St_\alpha\,\mathrm{d}\alpha$ is assumed.   These formal
properties greatly help understanding the generic behaviour of the
kinetic equation.  The
convergence issues are discussed later in \S\ref{sec:conv-coll-integr}.

First, (\ref{eq:kin sim}) conserves the total
wave energy and the two pseudomomentum components, i.e.,
\begin{equation}
  \label{eq:9}
  \dert{}\int (1,s_\alpha,r_\alpha)\,e_\alpha\,\mathrm{d}\alpha
  = 0.
\end{equation}
This is seen for the energy by rewriting the integrand of  (\ref{eq:kin sim}) as 
\begin{equation}
  \label{eq:10}
  \oma\,\Gamma^2
  \left(\frac{\oma}{\ea}+\frac{\omb}{\eb}+\frac{\omc}{\ec}\right)\ea\eb\ec, 
\end{equation}
which makes obvious that the integrand is $\oma$ times a
remainder that is totally symmetric in $(\alpha,\beta,\gamma)$.
After integration over $\alpha$ all the greek symbols can now be
freely relabelled and we can therefore rewrite the net integral
as an average over three copies of the original integral obtained
from the cyclic permutations of $(\alpha,\beta,\gamma)$.
Formally, these three integrals can then be combined to a single
integral with $\oma$ replaced by $(\oma+\omb+\omc)/3$.  This is
however zero on the resonant manifold, which proves energy
conservation.  The same argument also covers the pseudomomentum
components, e.g., $s_\alpha\oma=k_\alpha$ and
$(k_\alpha+k_\beta+k_\gamma)$ is also zero on the resonant
manifold.  This is an important standard argument in wave kinetic
theory.  These conservation laws persist after truncation of the
kinetic equation to a finite range of wavenumbers for a numerical
simulation.  This works provided only triads where all members
fall within the truncated range are considered.  This is a
slightly weaker form of the well known fact that the direct
numerical simulations of a Galerkin-truncated Boussinesq system
maintains the exact conservation of quadratic invariants.

Second, (\ref{eq:kin sim}) admits an irreversible entropy law of the form
\begin{equation}
  \label{eq:11}
  H = \int \ln e_\alpha\,\mathrm{d}\alpha\thf \dert{H}=\int \frac{St_\alpha}{\ea}\,\mathrm{d}\alpha
  \geq 0.
\end{equation}
Notably, the entropy density here is $\ln\ea$ and not $\ea\ln\ea$,
which means that the entropy law is restricted to strictly
positive distributions $\ea>0$.  The validity of (\ref{eq:11})
follows from dividing (\ref{eq:kin sim}) by $\ea$, integrating
over $\alpha$, averaging over the cyclic
permutations, and recombining into a single integral.  This replaces
(\ref{eq:10}) with the non-negative expression
\begin{equation}
  \label{eq:12}
  \frac{1}{3} \left(\frac{\oma}{\ea}+\frac{\omb}{\eb}+\frac{\omc}{\ec}\right)^2\Gamma^2\ea\eb\ec,
\end{equation}
which proves (\ref{eq:11}).  Consistent with the interpretation
of the kinetic equation as a nonlinear diffusion equation, there is a natural preponderance for the
collision integral to be positive in regions of small $\ea$ and
negative in regions of comparatively larger $\ea$.   Indeed,
if $\ea=0$ then (\ref{eq:kin sim}) has only a positive term
proportional to $\oma^2$ in the integrand, thus preserving $\ea\geq0$.    

Third, the distributions $\ea=(A+Bs_\alpha+Cr_\alpha)^{-1}$ with
constants $(A,B,C)$ are trivial steady states of (\ref{eq:kin
  sim}) because they make the integrand (\ref{eq:10}) vanish
on the resonant manifold.  These statistical equilibrium states are also
maximum entropy states for fixed energy and pseudomomentum, so
they are natural asymptotic end states for numerical
initial-value problems in a finite wavenumber domain.

Fourth, if the total wavenumber domain is decomposed into two
subdomains 
$\cA$ and $\cB$ that are mutually exclusive and jointly
exhaustive then the nonlinear energy exchange from $\cA$ to $\cB$
can be 
quantified unambiguously by $\int_\cB
St_\alpha\,\mathrm{d}\alpha$.  In particular,  
\begin{equation}\label{eq:Pi0}
    \Pi_0(K,t) =  \int\limits_{K_\alpha> K} St_\alpha\, K_\alpha \mathrm{d}{K}_\alpha\mathrm{d}\theta_\alpha
\end{equation}
\addOBB{is the  radial flux of energy from the interior
  of the circle $|\mathbf{k}_\alpha|=K$ to its exterior.} This holds whether or not the spectrum is isotropic
or in a steady state.

\subsection{Convergence of the collision integral}
\label{sec:conv-coll-integr}

Here we consider how the convergence of $\Sta$ and 
$\int\Sta\,\mathrm{d}\alpha$ depends on $e(\bk)$ (we omit the
time dependence, which is not relevant here). We fix $\bka\neq0$ and consider the convergence of $\Sta$. As
$\oma>0$ the resonance condition $\Omega\equiv\oma+\omb+\omc=0$
implies that at least one of $(\omb,\omc)$ is negative, so either
$\sigma_\beta=\sigma_\gamma=-1$ or $\sigma_\beta=-\sigma_\gamma$.
The delta
function in wavenumbers means that integration over
$\mathrm{d}\gamma$ merely sets $\bkc=-(\bka+\bkb)$ everywhere and thereafter
$\Sta$ amounts to a two-dimensional integral over $\bkb$.
Generically, the resonance condition $\Oma=0$ then selects a line
in the $\bkb$-plane and $\Sta$ reduces to an integral along this
line.  Convergence issues may then arise because of the frequency
delta function and/or in the limits $K_\beta\to\infty$ and
$K_\beta\to0$.

We will be most
interested in singular distributions of the form
\begin{equation}
  \label{eq:13}
  e(\bk) = e(K,\theta) = C_0 K^{-a} g(\theta)
\end{equation}
with Kolmogorov constant $C_0>0$, power law exponent $-a$, and a bounded
function $g(\theta)\geq0$ that captures the anisotropy.  Notably,
(\ref{eq:13}) includes monomial power
laws in the components of $\bka$, but it also includes many other
functions beyond that.

\subsubsection{Convergence of frequency delta function}
\label{sec:conv-freq-delta}

The argument of the
frequency delta function is
\begin{equation}
  \label{eq:14}
  \Omega(\bkb) = \oma +\sigma_\beta\omi(\bkb) +\sigma_\gamma \omi(-(\bka+\bkb))
\end{equation}
and a sufficient condition for convergence is that $\Omega=0$ implies
$|\grad_{\bkb}\Oma|\neq0$.   Depending on the sign of
$\sigma_\beta\sigma_\gamma$ this may fail if the group velocity
$\bcg=\grad_\bk\omi\neq0$ at $\bkb$ is
equal to plus or minus the group velocity at $\bkc$.  Now, it follows
from the dispersion relation (\ref{eq:disp}) that $\bcg(\bk)$ is
proportional to $1/K$ and directed along concentric circles
centred at $\bk=0$.  Specifically, $\bcg=0$ on both coordinate
axes and $\bcg$ points from the frequency minimum $f$ on the vertical axis $
k=0$ to the frequency maximum $N$ on the horizontal
axis $m=0$.     Hence $\bcg(\bk)\neq0$ can be inverted and
thus $\bcg(\bkb)=\pm\bcg(\bkc)$ implies
$\bkb=\pm\bkc$.  But then 
$(\bka,\bkb,\bkc)$ are collinear and share the same nonzero frequency in
magnitude, which means resonance is impossible.   

The remaining special
case where $\bcg(\bkb)=\bcg(\bkc)=0$ can occur only at the devilish
resonant triad
\begin{equation}
  \label{eq:15}
  (\oma,\omb,\omc) = (N-f,f,-N)
\end{equation}
with $\bkb=(0,-m_\alpha)$, $\bkc=(-k_\alpha,0)$, and $\omi(k_\alpha,m_\alpha)=N-f$.  
However, for this triad the second derivative matrix of
$\Oma(k_\beta,m_\beta)$ is positive definite, which
means $\Oma=0$ is an isolated zero and local minimum of $\Oma$.
In this special case the sufficient condition is not satisfied
but the delta function still converges, as can be verified  in the
example $\delta(x^2+y^2)$ integrated over the $xy$-plane.  The
conclusion is that there are no convergence problems for finite $K_\beta$ stemming
from the frequency delta function.

\subsubsection{Convergence at large or small $K_\beta$}
\label{sec:convergence-at-large}

We consider convergence as $K_\beta\to\infty$ at fixed
$\bka\neq0$.  In this limit $\bkc\approx-\bkb$ and therefore
$|\omb|\approx|\omc|$, which means resonance is possible only if
$\oma>2f$ and $(\omb,\omc)$ are both negative and approximately
equal to $-\oma/2$.  This triad is of parametric subharmonic
resonance type \citep{mccomas1977resonant}.  We first consider how $\Gamma$ in (\ref{eq:16})
scales with $K_\beta\gg1$. The pre-factor is bounded by
$1/K_\beta$, all the frequencies are proportional to
$\pm\oma$, the total slowness sum is bounded by $K_\alpha$, and the
slowness differences are bounded by $K_\beta$, so $\Gamma\sim
K_\alpha$ remains bounded in this limit.  For decaying
distributions like (\ref{eq:13}) we have that $\ea$ is
much larger than $\eb$ or $\ec$ so the term proportional
to $\eb\ec$ is negligible  in
(\ref{eq:kin sim}) compared to $\ea\eb\sim K_\beta^{-a}$.
Finally, the frequency delta function scales as $K_\beta$ because
the group velocities decay as $1/K_\beta$.  This leaves a line
integral proportional to $K_\beta^{1-a}\mathrm{d}K_\beta$, which
converges as $K_\beta\to\infty$ if $a>2$.

Conversely, if $K_\beta\to0$ at fixed
$\bka\neq0$ then $\bkc\approx-\bka$ and therefore
$|\omc|\approx\oma$.  Resonance then implies $\omc\approx\oma$
and $\omb\approx-2\oma$, which is possible if
$\oma<N/2$.  This is also a parametric subharmonic instability
triad.   Repeating the steps from before now leads to
$\Gamma^2\sim K_\beta^2$, a dominant energy term $\ea\eb\sim
K_\beta^{-a}$, and a delta function scaling with $K_\beta$
because of the dominant $|\bcg(\bkb)|\propto1/K_\beta$. This leaves a line
integral proportional to $K_\beta^{3-a}\mathrm{d}K_\beta$, which
converges as $K_\beta\to0$ if $a<4$.

The overall conclusion is that the collision integral $\Sta$ converges
at fixed $\bka\neq0$ 
for the class of singular distributions (\ref{eq:13}) with
\begin{equation}
  \label{eq:17}
2<a<4.  
\end{equation}
We note that in deriving \eqref{eq:17} we made frequent use of the strict bound $|\omega|\geq f>0$.
Convergence of $\Sta$ for (\ref{eq:13}) does not imply convergence of $\int\Sta\,\mathrm{d}\alpha$.  
In other words, for such singular
distributions global energy conservation cannot be assumed in its
simplest form (\ref{eq:9}), consistent with the infinite
energy content near the origin implied by (\ref{eq:17}).

\subsubsection{Self-similar collision integral}
\label{sec:self-simil-coll}

The power law distribution (\ref{eq:13}) makes the entire
integrand of $\Sta$ a homogeneous function of wavenumbers, which
allows extracting an exact self-similar form of the collision
integral.  Substituting (\ref{eq:13}) into \eqref{eq:kin
  sim}, rescaling $(\bkb,\bkc)$ by $|\bka|=K_\alpha$, and using
the wavenumber homogeneity of all factors in the integrand (degree two
for $\Gamma^2$, degree zero for the frequency delta function,
degree $-2a$ for $\eb\ec$, and degree two for $\mathrm{d}\bkb$)
then results in
\begin{equation}\label{eq:homcoll1}
  St_\alpha = St(K_\alpha,\theta_\alpha) =
  K_\alpha^{4-2a}\,\,St(1,\theta_\alpha).
   \end{equation}
Here $St(1,\theta_\alpha)$ is the collision integral evaluated on the unit circle
 $\bk_\alpha=(\cos\theta_\alpha,\sin\theta_\alpha)$, which depends
 implicitly on the exponent $a$ and on the angular dependence
 $g(\cdot)$ in (\ref{eq:13}).  We have already shown that $\Sta$ exists for
 $K_\alpha=1$ and therefore (\ref{eq:homcoll1}) establishes
 $\Sta$ for all finite nonzero $K_\alpha$.  Notably, 
$\int\Sta\mathrm{d}\alpha$ does not exist due to divergence
 either at zero ($a>3$),  at infinity ($a<3$), or both ($a=3$).

 We note in passing a peculiar finding, namely that if $a=4$ then
 (\ref{eq:kin sim}, \ref{eq:13}, \ref{eq:homcoll1}) formally
 admit time-dependent separable solutions of the form
 $e(\bk,t)=K^{-4}g(\theta,t)$.  It is unclear whether these
 solutions have physical meaning, and they occur at the boundary
 of the convergence interval (\ref{eq:17}), but they certainly
 provide a non-trivial test for numerical simulations of the
 kinetic equation.

\subsection{Turbulent spectrum}
\label{sec:turbulent-spectrum}

We turn to the main task of finding a turbulent spectrum of the
form (\ref{eq:13}) that corresponds to a steady state of the
kinetic equation with a nonzero spectral energy flux from small
to large wavenumbers.  In principle, to maintain such a steady
state requires forcing at small wavenumbers and damping at large
wavenumbers.  This would be the standard setting for a numerical
simulation and (\ref{eq:13}) would be the expected shape of the
spectrum in the inertial range far away from both the forcing and
the dissipation.  Actually, for the interval
(\ref{eq:17}) any region of wavenumber space that excludes a
patch around the origin contains finite wave energy so the
inertial range may extend to infinite wavenumber and there is no
need to consider large wavenumber dissipation explicitly (i.e.,
the inertial range has `finite capacity' in the terminology of
wave turbulence theory).  Hence (\ref{eq:kin sim}) is replaced by
\begin{equation}
  \label{eq:18}
   \partial_t \ea = St_\alpha + F_\alpha
\end{equation}
where the forcing function $F_\alpha\geq0$ is supported in a
small neighborhood of the origin, e.g., $F_\alpha=0$ if
$K_\alpha>\epsilon$ for some $\epsilon\ll1$, and we may assume
that the forcing amplitude is scaled such that net  input rate
$\int F_\alpha\,\mathrm{d}\alpha$ is independent of $\eps$. 

In a steady state it follows that $\Sta=-F_\alpha$, i.e., the collision integral is equal to zero
everywhere except in the forcing region, where it precisely
cancels the forcing.  Then the radial energy flux $\Pi_0(K)$ from
(\ref{eq:Pi0}) is constant and equal to the net input rate for
all $K>\eps$.  The question is whether it is possible to
find a steady-state spectrum that is of the form (\ref{eq:13}) in
the far field 
$K_\alpha\gg\eps$ and can be modified in and near the forcing
region to yield a finite energy content and also to satisfy
$\Sta=-F_\alpha\leq0$ there.  We present three different
arguments 
that strongly suggest that there is only a single power law that
is a candidate for such a solution.

The first result in this direction
was given in \cite{shavit2024sign}, which extended standard arguments
   for the collision integral from isotropic wave turbulence \cite{ZLF} to the present
   strongly anisotropic case.   Specifically, the
   important homogeneity methods
   developed in \cite{balk2000kolmogorov} were used to show that for the special
   value
   \begin{equation}
     \label{eq:19}
     a_0=3
   \end{equation}
   the angle average
\begin{equation}
  \label{eq:20}
  \overline{St(1,\theta_\alpha)} = \frac{1}{2\pi}\int_0^{2\pi}
  St(1,\theta_\alpha)\,\mathrm{d}\theta_\alpha 
\end{equation}
vanishes, provided
   $St(1,\theta_\alpha)$ exists in the first place, which we know to be
   true here.  Notably, this result did not require finding the
   angular dependence of the spectrum, in fact if $a=3$ then the
   angle average was shown to be zero for \textsl{any} bounded choice of $g(\cdot)$.  More
   generally, with the same construction one finds that
\begin{equation}\label{eq:hom_degree}
    a_0=a_V+d-a_\omega/2,
\end{equation}
where $a_V=1$ is the homogenity  degree of the interaction
coefficients, $d=2$ is the number of spatial dimensions, and
$a_\omega=0$ is the homogenity degree of the frequency
\citep{shavit2024sign}.

A second heuristic argument for (\ref{eq:19}) can be obtained by
computing the radial energy flux \eqref{eq:Pi0} based on the self-similar collision integral, which converges at large wavenumbers if $a>3$ \citep{shavit2025turbulent}.  This is motivated by
(\ref{eq:18}) and the expected modification of the singular
behaviour of $\ea$ near the origin but not in the far field.   This leads to
\begin{equation}\label{eq:fluxang} 
 \Pi_0(K)=-\frac{K^{6-2a}}{6-2a}
 2\pi C_0^2\,\, \overline{St_{\theta}},
\end{equation}
where $St_{\theta}=St(1,\theta_\alpha)/C_0^2$.  This shows that for a
steady state with $St_\theta=0$ the self-similar radial flux is
zero except possibly at the exceptional value $a=3$, where a
nonzero limit might be extracted from the indeterminate form in
(\ref{eq:fluxang}).  Again, this argument does not involve the
angular dependence $g(\cdot)$ of the spectrum.

A third heuristic argument comes from total energy conservation,
which implies that if $\int\Sta\,\mathrm{d}\alpha$ converges then it
must be zero.  The angle average  $\overline{\Sta}$ in (\ref{eq:homcoll1}) has the
same sign for all $K_\alpha$ so taken at face value this would imply that
 $\overline{St_\theta}=0$ for any $g(\cdot)$.  However, as already
 stated, the  total integral of (\ref{eq:homcoll1}) either
diverges algebraically 
at the origin (if $a>3$) or at infinity (if $a<3$), so no conclusion
about $\overline{St_\theta}$ can be inferred in those cases.  The two ranges
touch at $a=3$ and then the divergences are merely logarithmic.
This suggests again that $a=3$ is special and that then 
$\overline{St_\theta}=0$ follows from energy conservation.

Taken together these arguments strongly suggest that $a=3$ is the
only relevant power law exponent for a turbulent spectrum.
Incidentally, this value is precisely halfway in the convergence
interval $2<a<4$, which apparently is a relatively frequent
occurence in isotropic wave turbulence \citep{ZLF}, and one that
seemingly carries over to the present strongly anisotropic case.

Finally, a necessary condition for this construction to work is that the
spectrum (\ref{eq:13}) can be substantially modified in the forcing region
$K_\alpha<\eps$ with negligible impact on $\Sta$ in the
far-field region $K_\alpha\gg\eps$.   This condition is  demonstrably
true in the present case, based on the considerations already
presented in \S\ref{sec:convergence-at-large} for small
$K_\beta$.  Specifically, if  $K_\alpha=O(1)$ then the only triads 
linking to $\bkb$ in the forcing region $K_\beta<\eps$ are parametric
subharmonic instability triads with $\omb=-2\oma$.  As pointed
out before, there are no such triads if $\oma>N/2$, so actually that entire
high-frequency sector $\oma>N/2$ is strictly disconnected from the forcing
region  for small enough $\eps$.  If $\oma<N/2$ then triad
members  in
the forcing region do exist and fall 
along the line $|\omb|=2\oma$, but as shown their net contribution  
integrates only  to $\int_0^\eps
K_\beta^{3-a}\mathrm{d}K_\beta=O(\eps)$, which is negligible.  This bound is
based on what is arguably the worst case: extending the
singular power law spectrum all the way to the origin in
wavenumber space.   The true spectrum would actually be much
smaller near the origin, which strengthens the conclusion.   Hence the interactions
with the small forcing region are indeed negligible in the far
field, which allows an inertial range of the form (\ref{eq:13})
to be a solution there.

\section{The angular part of the spectrum}
\label{sec:angpart}

The angular part $g(\theta)$ of the steady turbulent spectrum
(\ref{eq:13}) must be found numerically, which requires
parametrizing the resonant manifold in the collision integral and
then evaluating it numerically to ensure $St_\theta=0$ for all
angles.
This non-trivial task is
detailed in appendix \S\ref{sec:app}. In practice, the search has to be
restricted to find optimal parameters in some assumed functional
shape for $g(\theta)$, which involves some inspired guesswork.  

In \cite{shavit2025turbulent} this task was undertaken for the
non-rotating case $f=0$, but the presence of zero-frequency waves
along the vertical axis in wavenumber space led to singular
behaviour of the collision integral there.  This was resolved in
an \textsl{ad hoc} fashion by
removing a small sector of frequencies $\delta>0$ around the vertical axis,
which produced the lower bound $|\omega|\geq\delta$ and thus allowed
convergence of the collision integral.  A functional shape 
\begin{equation}
  \label{eq:angturbspec}
  g(\theta) = {\omi(\theta)^{2b}}
\end{equation}
where $\omi(\theta)$ is the positive root of (\ref{eq:omega}) was
assumed.  Notably, a spectrum of the form (\ref{eq:angturbspec})
has additional symmetries such as $g(\theta+\pi)=g(\theta)$ that imply that there is no net wave
propagation in any direction and consequently both components of
the net pseudomomentum are zero.  The optimal exponent $b$ was found by computing
$St_\theta$ for many different test values of $\theta$.  This
numerical process was then repeated with a reduced value for
$\delta$ and very strong numerical evidence was obtained for
convergence $b\to-1$ as $\delta\to0$.   This led to the
regularized turbulent spectrum $e=C_0 K^{-3}/\omega^2=(C_0/N^2) K^{-3}/\cos^2\theta$.  As
pointed out in the introduction, this spectrum agrees
with the two-dimensional GM spectrum for large
$\omega/f$.  However, at small frequencies this spectrum has
non-integrable singularities along the entire vertical axis in
wavenumber space, so $\delta$ can be arbitrarily small but not
zero.

\begin{figure}
\includegraphics[scale=0.5]{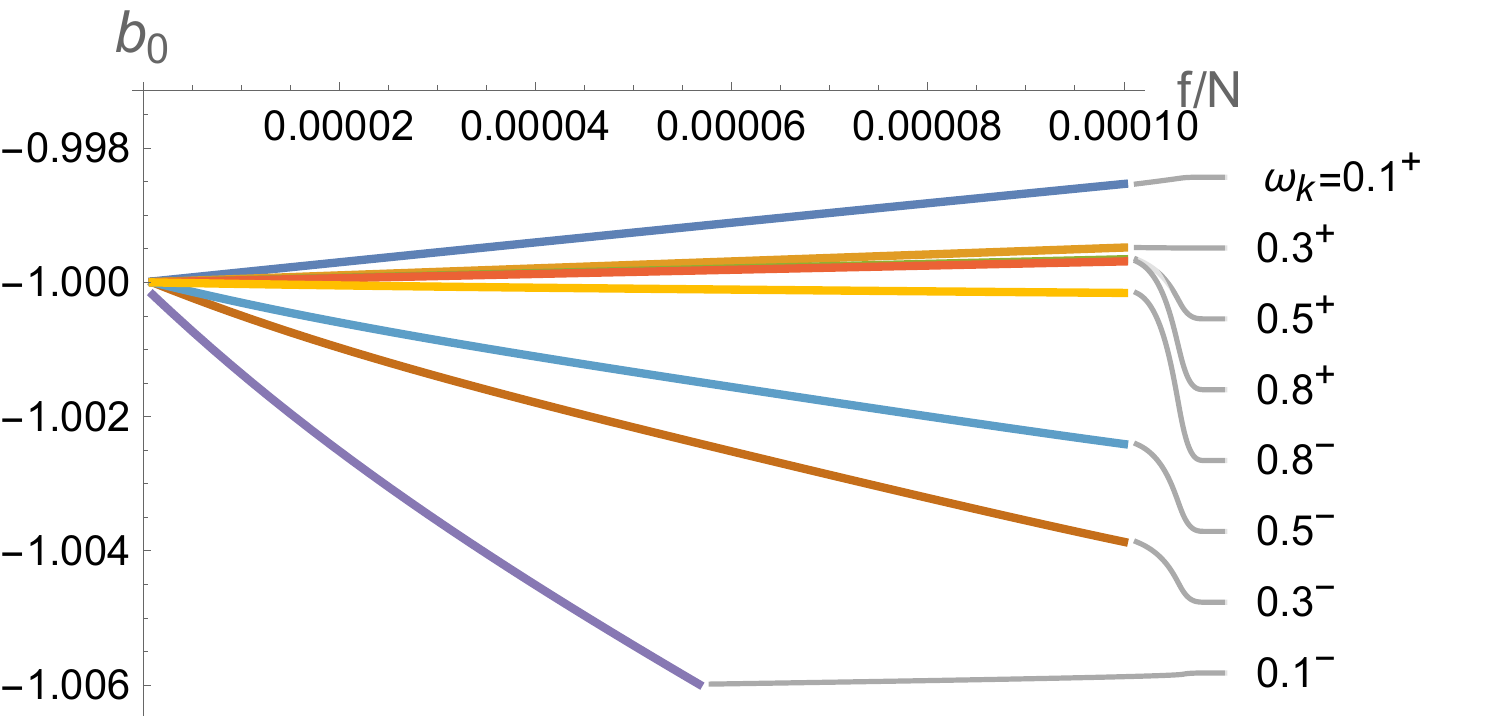} 
\caption{\label{fig:b_f}
We use (\ref{eq:13}) with $a=3$ and search for convergence of the power law exponent $b\to b_0$ in  $g(\theta)=\omega(\theta)^{2b}$ as $f/N\to0$. For numerical convenience, we split  $St_\theta$ into parts $St^\pm_\theta$ that converge from above and below.  
For various test frequencies $\omega_k$ we determine numerically the values $b^{\pm}$ for which $St_{\theta_k}^{\pm}$ vanishes; these are plotted as functions of  $f/N$. We find that both $b^\pm$ approach $b_0=-1$ as rotation diminishes.
}
\end{figure}

For the rotating case $f\neq0$ we retained the same functional
shape (\ref{eq:angturbspec}) and followed the same numerical
method, which however required a much more complicated
parametrization of the rotating resonant manifold.\footnote{That the rotating resonant manifold is a complicated analytical object is hinted at by the fact that if $f\to N/2$ the resonant manifold disappears.}  Our idea was
that any nonzero $f/N$ would act as a natural physical
regularizer that bounds the frequencies away from zero and thus
prevents the singularities we encountered in the non-rotating
case.  With this limited parametric search strategy we were not able to find a convincing turbulent solution for
finite values of $f/N$, but very convincing numerical evidence was
obtained that $b=-1$ emerges again in the limit of infinitesimal
rotation $f/N\to0$.  Our process fixed $a=3$ and for decreasing
values of $f/N$ sought the exponent $b$ that makes the 
collision integral vanish at a set of different test frequencies $\omega(\theta)$. Of course, for
there to be a solution the same exponent $b$ should work for all
test frequencies.  We found that for finite $f/N$ the values
of $b$ were different for different angles $\theta$, but there
was clear evidence of convergence to $b=-1$ across
all angles in the
limit $f/N\to0$, see figure
\ref{fig:b_f}.   This yields the limiting turbulent spectrum
\begin{equation}\label{eq:vanishspec}
    \lim_{f/N\rightarrow 0} e(K,\theta) = C_0\,K^{-3}\omega(\theta)^{-2}.
\end{equation}
For any nonzero $f/N$ this spectrum is integrable across
vertical axis, which is crucial for its physical relevance.  This is reminiscent of the
regularized singularity encountered in the non-rotating case.     

Technically, infinitesimal rotation can be understood as a
regularizer that separates low-frequency waves from zero-frequency
balanced modes. At $f/N=0$, these
frequencies touch on the line $\theta=\pi/2$ and fundamental assumptions used
in the derivation of the kinetic equation break down. For example, the
assumption of Gaussianity fails, and
$\left<Z_{(+,0,m)}Z_{(-,0,m)}^{*}\right>\neq0$, see
\citep{shavit2025turbulent}. From this point of view, working with
infinitesimal rotation generalizes our previous approach, in
which we introduced an ad-hoc regularizer around slow modes, and
found the turbulent spectrum in the limit as the regularizer width
goes to zero. Here we showed explicitly that rotation acts as a
physical regularizer around slow modes, so that the turbulent
spectrum emerges (\ref{eq:vanishspec}) naturally.
We note again that transforming $C_0\,K^{-3}\omega(\theta)^{-2}$ to a density in $(\omega,m)$-space by using the full dispersion relation yields \eqref{eq:22}, which is the GM spectrum \eqref{eq:GMspec} for all frequencies.

\section{Concluding remarks}
\label{sec:concl}

We have provided a first-principles route from the two-dimensional rotating–stratified Boussinesq equations to the observed internal-wave spectrum. The key ingredients are: an explicit restriction to the wave manifold by imposing the linear PV constraint and thereby eliminating balanced modes; a resonant kinetic equation whose interaction coefficients factor as $V_\alpha = \omega_\alpha\,\Gamma$ on the resonant manifold, making the conservation of energy and pseudomomentum completely transparent; and the homogeneity of both the dispersion relation and the interaction coefficients, which selects a unique flux-carrying steady state of the form $e(K,\theta)=C_0 K^{-3} g(\theta)$. Within this framework narrowband forcing at large scales feeds a finite-capacity inertial range in which the details of the forcing and small-scale dissipation become irrelevant, and the spectrum is instead determined by the structure of the kinetic equation itself.

Mapping this homogeneous solution into observational variables $(\omega,m)$ shows that the turbulent spectrum is necessarily separable: the measure transforms into a product of a vertical factor $m^{-2}$ and a frequency factor $\omega^{-2}$ multiplied by a Jacobian factor $\omega/\sqrt{\omega^2 - f^2}$, which yields the GM spectrum for the whole frequency range.   Notably, any different angle dependence $g(\theta)\neq \omega^{-2}$ would still lead to a separable spectral density in $(\omega,m)$-space. 

The weak but nonzero rotation regularizes the problem by eliminating the interactions with slow modes that were singular in the strictly non-rotating case. From this perspective, the weak-rotation limit is not a minor technical assumption, but the mechanism that both selects the GM law and controls how the spectrum organizes near the inertial frequency.

These results have several conceptual implications. First, the observed separability of the internal-wave spectrum is not an empirical coincidence: it follows directly from homogeneity and the existence of a turbulent, flux-carrying solution of the kinetic equation. Second, rotation acts as a natural regulator of interactions with slow modes, cleanly separating internal-wave dynamics from balanced motions and clarifying how energy exchange is organized near the inertial frequency. Third, extending the homogeneity analysis to three dimensions suggests  a flux-carrying solution $e = C_0 K^{-4} g(\theta,\phi)$ in polar coordinates. Only the exponent $a=4$ is known here.  By assuming horizontal isotropy (or by averaging over the azimuthal angle $\phi$) this spectrum again becomes separable in $(\omega,m)$-space, with the polar Jacobian factor smoothing the inertial singularity that appeared in two dimensions.

\textit{\textbf{Acknowledgments}}.
 %\addOB{The referees' comments significantly improved the original manuscript.} 
This work was supported by the Simons Foundation and the Simons Collaboration on Wave Turbulence.
  %The numerical study was made
  %possible thanks to New York University's Greene computing
  %cluster facility.
OB acknowledges additional financial support under  NSF grant
DMS-2406767. MS acknowledges additional financial support from
the Schmidt Futures Foundation. %and the Israeli CHE.

\appendix
\label{sec:app}

\section{The resonant manifold and integration of the collision integral}
Remind that the kinetic equation is
\begin{align}\label{eq:kin_appen}
    \dot{e_{\alpha}}=&St_{\alpha}\\=&\pi\int d\beta d\gamma\omega_{\alpha}\Gamma_{\alpha\beta\gamma}^{2}\left(\omega_{\beta}e_{\alpha}e_{\gamma}+\omega_{\gamma}e_{\beta}e_{\alpha}+\omega_{\alpha}e_{\beta}e_{\gamma}\right)\delta\left(\mathbf{k}_{\alpha}+\mathbf{k}_\beta+\mathbf{k}_\gamma\right)\delta\left(\omega_\alpha+\omega_{\beta}+\omega\gamma\right).
\end{align}
In polar coordinates the resonant manifold is given by the following three equations 
\begin{align}
K_{\alpha}\sin\theta_{\alpha}+K_{\beta}\sin\theta_{\beta}+K_{\gamma}\sin\theta_{\gamma}&=0,\\K_{\alpha}\cos\theta_{\alpha}+K_{\beta}\cos\theta_{\beta}+K_{\gamma}\cos\theta_{\gamma}&=0,\\\omega_{\alpha}+\omega_{\beta}+\omega_{\gamma}&=0. 
\end{align}
The third equation implies that there are no interactions among waves of the same branch. First two equations fix
\begin{align}\label{K's}
K_{\beta}&=K_{\alpha}\frac{\cos\theta_{\alpha}\sin\theta_{\gamma}-\sin\theta_{\alpha}\cos\theta_{\gamma}}{\cos\theta_{\gamma}\sin\theta_{\beta}-\cos\theta_{\beta}\sin\theta_{\gamma}}=K_{\alpha}\frac{\sin\left(\theta_{\gamma}-\theta_{\alpha}\right)}{\sin\left(\theta_{\beta}-\theta_{\gamma}\right)}\\K_{\gamma}&=-K_{\alpha}\frac{\cos\theta_{\alpha}\sin\theta_{\beta}-\sin\theta_{\alpha}\cos\theta_{\beta}}{\cos\theta_{\gamma}\sin\theta_{\beta}-\cos\theta_{\beta}\sin\theta_{\gamma}}=-K_{\alpha}\frac{\sin\left(\theta_{\beta}-\theta_{\alpha}\right)}{\sin\left(\theta_{\beta}-\theta_{\gamma}\right)}=K_{\alpha}\frac{\sin\left(\theta_{\alpha}-\theta_{\beta}\right)}{\sin\left(\theta_{\beta}-\theta_{\gamma}\right)}
\end{align}
The constraint $0<K_{\beta},K_{\gamma}$ limits the angular integration $0\leq\theta_{\beta},\theta_{\gamma}\leq2\pi$ to 
\begin{align}\label{angular_limits}
1:\sin\left(\theta_{\beta}-\theta_{\gamma}\right)>0&:\sin\left(\theta_{\alpha}-\theta_{\beta}\right)>0\&\sin\left(\theta_{\gamma}-\theta_{\alpha}\right)>0\\2:\sin\left(\theta_{\beta}-\theta_{\gamma}\right)<0&:\sin\left(\theta_{\alpha}-\theta_{\beta}\right)<0\&\sin\left(\theta_{\gamma}-\theta_{\alpha}\right)<0
\end{align}
take $0\leq\theta_{\alpha}\leq\pi/2$, \eqref{angular_limits} gives
\begin{align}
    0\leq\theta_{\beta}-\theta_{\gamma}\leq\pi&\,\,or\,\,-2\pi\leq\theta_{\beta}-\theta_{\gamma}\leq-\pi\\0\leq\theta_{\alpha}-\theta_{\beta}\leq\pi&\,\,or\,\,-2\pi\leq\theta_{\alpha}-\theta_{\beta}\leq-\pi\\0\leq\theta_{\gamma}-\theta_{\alpha}\leq\pi&
\end{align}
and (2) in \eqref{angular_limits}  gives 
\begin{align}
    \pi\leq\theta_{\beta}-\theta_{\gamma}\leq2\pi&\,\,or\,\,-\pi\leq\theta_{\beta}-\theta_{\gamma}\leq0\\&\,\,\,\,\,\,-\pi\leq\theta_{\alpha}-\theta_{\beta}\leq0\\\pi+\theta_{\alpha}\leq\theta_{\gamma}\leq2\pi&\,\,or\,\,0\leq\theta_{\gamma}\leq\theta_{\alpha}
\end{align}
\subsection{$\theta_\alpha=0, \omega_\alpha=N$}
We explain in detail here the parameterization of the resonant manifold for $\theta_\alpha=0$ and take $\omega_\alpha=N$, the case of general $\theta_\alpha$ is included in a Mathematica notebook. 

The integration limits given by (1) in \eqref{angular_limits} 
\begin{align}\label{angular_limits_N}
    \pi\leq\theta_{\beta}&\leq\min\left\{ \pi+\theta_{\gamma},2\pi\right\}, \\0\leq\theta_{\gamma}&\leq\pi.
\end{align}
(2) is symmetric wrt the interchange $\beta\leftrightarrow\gamma$. 
To solve the resonant condition $\omega_\alpha+\omega_\beta+\omega_\gamma=0$ we change the integration from angles to positive frequencies. Let $\beta=\left(\pm,\mathbf{p}\right)$ and denote 
\begin{equation}
\omega_{p}=\sqrt{N^{2}\cos^{2}\theta_{k}+f^{2}\sin^{2}\theta_{k}}
\end{equation}
then
\begin{equation}
\int_{\theta_{1}}^{\theta_{2}}d\theta_{\beta}=\int_{\min\left\{ \omega_{2},\omega_{1}\right\} }^{\max\left\{ \omega_{2},\omega_{1}\right\} }\frac{\omega_{p}d\omega_{p}}{\left(N^{2}-f^{2}\right)\left|\cos\theta_{p}\right|\left|\sin\theta_{p}\right|}=\int_{\min\left\{ \omega_{2},\omega_{1}\right\} }^{\max\left\{ \omega_{2},\omega_{1}\right\} }\frac{\omega_{p}d\omega_{p}}{\sqrt{\omega_{p}^{2}-f^{2}}\sqrt{N^{2}-\omega_{p}^{2}}}.
\end{equation}
\eqref{angular_limits_N} gives three regions 
\begin{align}\nonumber
B1:&0\leq\theta_{\gamma}\leq\pi/2,\pi\leq\theta_{\beta}\leq\pi+\theta_{\gamma},
\\B2:&\pi/2\leq\theta_{\gamma}\leq\pi,\pi\leq\theta_{\beta}\leq\pi+\pi/2, \nonumber
\\B3:&\pi/2\leq\theta_{\gamma}\leq\pi,\pi+\pi/2\leq\theta_{\beta}\leq\pi+\theta_{\gamma\nonumber }
\end{align}
Let $\beta=\left(\pm,\mathbf{p}\right)$ and $\gamma=\left(\pm,\mathbf{q}\right)$. B1,B2,B3 have solutions only on $\left(+,-,-\right)$ which sets $\omega_{p}=\omega_{k}-\omega_{q}$ and the remaining integration 
\begin{align}\nonumber
    B1:&f<\omega_{q}<0.5\omega_{k}=0.5N,
    \\B2:&f<\omega_{q}<\omega_{k}-f=N-f, \nonumber
    \\B3:&\omega_{k}/2<\omega_{q}<\omega_{k}-f=N-f.\nonumber
\end{align}
The collision integral, \eqref{eq:kin sim} is now reduced to the 1D integral  
\begin{equation}
    St_{k}=\pi\omega_{k}\int d\omega_{q}J\left(\omega_{q}\right)\Gamma_{\alpha\beta\gamma}^{2}\left(\omega_{k}e_{p}e_{q}-\omega_{p}e_{k}e_{q}-\omega_{q}e_{p}e_{q}\right),
\end{equation}
where 
\begin{align}
\Gamma_{\alpha\beta\gamma}^{2}&=\frac{\left(V_{\alpha}^{\beta\gamma}\right)^{2}}{\omega_{\alpha}^{2}}\propto\frac{\left(\cos\theta_{\beta}\sin\theta_{\gamma}-\sin\theta_{\beta}\cos\theta_{\gamma}\right)^{2}}{K_{\alpha}\omega_{\alpha}^{2}}\left(K_{\gamma}^{2}-K_{\beta}^{2}+f^{2}s_{\alpha}^{z}\left(s_{\gamma}^{z}-s_{\beta}^{z}\right)+N^{2}s_{\alpha}^{x}\left(s_{\gamma}^{x}-s_{\beta}^{x}\right)\right)\\&=\frac{\left(\cos\theta_{\beta}\sin\theta_{\gamma}-\sin\theta_{\beta}\cos\theta_{\gamma}\right)^{2}}{K_{\alpha}\omega_{\alpha}^{2}}\times\\&\left(K_{\gamma}^{2}-K_{\beta}^{2}+f^{2}K_{\alpha}\frac{\sin\theta_{\alpha}}{\omega_{k}}\left(K_{\gamma}\frac{\sin\theta_{\gamma}}{-\omega_{q}}-K_{\beta}\frac{\sin\theta_{\beta}}{-\omega_{p}}\right)+N^{2}K_{\alpha}\frac{\cos\theta_{\alpha}}{\omega_{k}}\left(K_{\gamma}\frac{\cos\theta_{\gamma}}{-\omega_{q}}-K_{\beta}\frac{\cos\theta_{\beta}}{-\omega_{p}}\right)\right)
\end{align}
and the Jacobian
\begin{equation}
    J\left(\omega_{q}\right)	=\frac{1}{\left|\cos\theta_{\gamma}\sin\theta_{\beta}-\cos\theta_{\beta}\sin\theta_{\gamma}\right|}\frac{\omega_{p}}{\sqrt{\omega_{p}^{2}-f^{2}}\sqrt{N^{2}-\omega_{p}^{2}}}\frac{\omega_{q}}{\sqrt{\omega_{q}^{2}-f^{2}}\sqrt{N^{2}-\omega_{q}^{2}}}
\end{equation}
with $\omega_p=\omega_k-\omega_q$, the absolute wave numbers $K_\beta,K_\gamma$ given by \eqref{K's} and on 
B1:
\begin{align}
    \cos\theta_{p}	=-\sqrt{\frac{\omega_{p}^{2}-f^{2}}{N^{2}-f^{2}}},
\,\,\sin\theta_{p}	=-\sqrt{\frac{N^{2}-\omega_{p}^{2}}{N^{2}-f^{2}}},
\,\,\cos\theta_{q}	=\sqrt{\frac{\omega_{q}^{2}-f^{2}}{N^{2}-f^{2}}},
\,\,\sin\theta_{q}	=\sqrt{\frac{N^{2}-\omega_{q}^{2}}{N^{2}-f^{2}}}
\end{align}
B2:
\begin{align}
    \cos\theta_{p}	=-\sqrt{\frac{\omega_{p}^{2}-f^{2}}{N^{2}-f^{2}}},
\,\,\sin\theta_{p}	=-\sqrt{\frac{N^{2}-\omega_{p}^{2}}{N^{2}-f^{2}}},
\,\,\cos\theta_{q}	=-\sqrt{\frac{\omega_{q}^{2}-f^{2}}{N^{2}-f^{2}}},
\,\,\sin\theta_{q}	=\sqrt{\frac{N^{2}-\omega_{q}^{2}}{N^{2}-f^{2}}}
\end{align}
B3:
\begin{align}
    \cos\theta_{p}	=\sqrt{\frac{\omega_{p}^{2}-f^{2}}{N^{2}-f^{2}}},
\,\,\sin\theta_{p}	=-\sqrt{\frac{N^{2}-\omega_{p}^{2}}{N^{2}-f^{2}}},
\,\,\cos\theta_{q}	=-\sqrt{\frac{\omega_{q}^{2}-f^{2}}{N^{2}-f^{2}}},
\,\,\sin\theta_{q}	=\sqrt{\frac{N^{2}-\omega_{q}^{2}}{N^{2}-f^{2}}}.
\end{align}
The 1D integration of the collision integral is done numerically and is included in the Mathematica notebook.

\bibliographystyle{jfm}
\bibliography{main}

\begin{thebibliography}{27}
\expandafter\ifx\csname natexlab\endcsname\relax\def\natexlab#1{#1}\fi
\def\au#1{#1} \def\ed#1{#1} \def\yr#1{#1}\def\at#1{#1}\def\jt#1{\textit{#1}} \def\bt#1{#1}\def\bvol#1{\textbf{#1}} \def\vol#1{#1} \def\pg#1{#1} \def\publ#1{#1}\def\arxiv#1{#1}\def\org#1{#1}\def\st#1{\textit{#1}}

\bibitem[Balk(2000)]{balk2000kolmogorov}
{\sc \au{Balk, AM}} \yr{2000}  \at{On the {K}olmogorov--{Z}akharov spectra of weak turbulence}.  \jt{Physica D: Nonlinear Phenomena}  \bvol{139}~(1-2),  \pg{137--157}.

\bibitem[B{\"u}hler {\em et~al.\/}(1999)B{\"u}hler, McIntyre \& Scinocca]{buhler1999shear}
{\sc \au{B{\"u}hler, Oliver}, \au{McIntyre, Michael~E} \& \au{Scinocca, John~F}} \yr{1999}  \at{On shear-generated gravity waves that reach the mesosphere. part i: Wave generation}.  \jt{Journal of the atmospheric sciences}  \bvol{56}~(21),  \pg{3749--3763}.

\bibitem[Dematteis \& Lvov(2023)]{dema23}
{\sc \au{Dematteis, Giovanni} \& \au{Lvov, Yuri~V.}} \yr{2023}  \at{The structure of energy fluxes in wave turbulence}.  \jt{Journal of Fluid Mechanics}  \bvol{954},  \pg{A30}.

\bibitem[Ferrari \& Wunsch(2009)]{ferrari2009ocean}
{\sc \au{Ferrari, Raffaele} \& \au{Wunsch, Carl}} \yr{2009}  \at{Ocean circulation kinetic energy: Reservoirs, sources, and sinks}.  \jt{Annual Review of Fluid Mechanics}  \bvol{41}~(1),  \pg{253--282}.

\bibitem[Galtier(2022)]{galtier2022physics}
{\sc \au{Galtier, S{\'e}bastien}} \yr{2022} {\em Physics of wave turbulence\/}.  \publ{Cambridge University Press}.

\bibitem[Garrett \& Munk(1979)]{garrett1979internal}
{\sc \au{Garrett, Christopher} \& \au{Munk, Walter}} \yr{1979}  \at{Internal waves in the ocean}.  \jt{Annual review of fluid mechanics}  \bvol{11}~(1),  \pg{339--369}.

\bibitem[Hasselmann(1966)]{hasselmann1966feynman}
{\sc \au{Hasselmann, Klaus}} \yr{1966}  \at{Feynman diagrams and interaction rules of wave-wave scattering processes}.  \jt{Reviews of Geophysics}  \bvol{4}~(1),  \pg{1--32}.

\bibitem[Labarre {\em et~al.\/}(2024{\natexlab{{\em a\/}}})Labarre, Augier, Krstulovic \& Nazarenko]{labarre2024internal}
{\sc \au{Labarre, Vincent}, \au{Augier, Pierre}, \au{Krstulovic, Giorgio} \& \au{Nazarenko, Sergey}} \yr{2024{\natexlab{{\em a\/}}}}  \at{Internal gravity waves in stratified flows with and without vortical modes}.  \jt{Physical Review Fluids}  \bvol{9}~(2),  \pg{024604}.

\bibitem[Labarre {\em et~al.\/}(2024{\natexlab{{\em b\/}}})Labarre, Lanchon, Cortet, Krstulovic \& Nazarenko]{labarre2024kinetics}
{\sc \au{Labarre, Vincent}, \au{Lanchon, Nicolas}, \au{Cortet, Pierre-Philippe}, \au{Krstulovic, Giorgio} \& \au{Nazarenko, Sergey}} \yr{2024{\natexlab{{\em b\/}}}}  \at{On the kinetics of internal gravity waves beyond the hydrostatic regime}.  \jt{Journal of Fluid Mechanics}  \bvol{998},  \pg{A17}.

\bibitem[Labarre \& Shavit(2024)]{labarre20242d}
{\sc \au{Labarre, Vincent} \& \au{Shavit, Michal}} \yr{2024}  \at{2d internal gravity wave turbulence}.  \jt{arXiv preprint arXiv:2412.20534} .

\bibitem[Lanchon \& Cortet(2023)]{lanchon2023energy}
{\sc \au{Lanchon, Nicolas} \& \au{Cortet, Pierre-Philippe}} \yr{2023}  \at{Energy spectra of nonlocal internal gravity wave turbulence}.  \jt{Physical Review Letters}  \bvol{131}~(26),  \pg{264001}.

\bibitem[Lvov {\em et~al.\/}(2010)Lvov, Polzin, Tabak \& Yokoyama]{lvov2010oceanic}
{\sc \au{Lvov, Yuri~V}, \au{Polzin, Kurt~L}, \au{Tabak, Esteban~G} \& \au{Yokoyama, Naoto}} \yr{2010}  \at{Oceanic internal-wave field: Theory of scale-invariant spectra}.  \jt{Journal of Physical Oceanography}  \bvol{40}~(12),  \pg{2605--2623}.

\bibitem[Lvov \& Tabak(2001)]{lvov2001hamiltonian}
{\sc \au{Lvov, Yuri~V} \& \au{Tabak, Esteban~G}} \yr{2001}  \at{Hamiltonian formalism and the garrett-munk spectrum of internal waves in the ocean}.  \jt{Physical review letters}  \bvol{87}~(16),  \pg{168501}.

\bibitem[MacKinnon {\em et~al.\/}(2017)MacKinnon, Zhao, Whalen, Waterhouse, Trossman, Sun, Laurent, Simmons, Polzin, Pinkel {\em et~al.\/}]{mackinnon2017climate}
{\sc \au{MacKinnon, Jennifer~A}, \au{Zhao, Zhongxiang}, \au{Whalen, Caitlin~B}, \au{Waterhouse, Amy~F}, \au{Trossman, David~S}, \au{Sun, Oliver~M}, \au{Laurent, Louis C~St}, \au{Simmons, Harper~L}, \au{Polzin, Kurt}, \au{Pinkel, Robert} \& \au{others}} \yr{2017}  \at{Climate process team on internal wave--driven ocean mixing}.  \jt{Bulletin of the American Meteorological Society}  \bvol{98}~(11),  \pg{2429--2454}.

\bibitem[McComas \& Bretherton(1977)]{mccomas1977resonant}
{\sc \au{McComas, C~Henry} \& \au{Bretherton, Francis~P}} \yr{1977}  \at{Resonant interaction of oceanic internal waves}.  \jt{Journal of Geophysical Research}  \bvol{82}~(9),  \pg{1397--1412}.

\bibitem[Munk(1981)]{munk1981internal}
{\sc \au{Munk, WH}} \yr{1981}  \at{Internal waves and small-scale processes}.  \jt{Evolution of physical oceanography} .

\bibitem[Nazarenko(2011)]{nazabook}
{\sc \au{Nazarenko, Sergey}} \yr{2011} {\em Wave turbulence\/}, ,  \vol{vol. 825}.  \publ{Springer Science \& Business Media}.

\bibitem[Olbers(1983)]{olbers1983models}
{\sc \au{Olbers, Dirk~J}} \yr{1983}  \at{Models of the oceanic internal wave field}.  \jt{Reviews of Geophysics}  \bvol{21}~(7),  \pg{1567--1606}.

\bibitem[Ripa(1981)]{ripa}
{\sc \au{Ripa, P}} \yr{1981}  \at{On the theory of nonlinear wave-wave interactions among geophysical waves}.  \jt{Journal of Fluid Mechanics}  \bvol{103},  \pg{87--115}.

\bibitem[Salmon(1988)]{salmon1988hamiltonian}
{\sc \au{Salmon, Rick}} \yr{1988}  \at{Hamiltonian fluid mechanics}.  \jt{Annual review of fluid mechanics}  \bvol{20}~(1),  \pg{225--256}.

\bibitem[Salmon(1998)]{salmon1998lectures}
{\sc \au{Salmon, Rick}} \yr{1998} {\em Lectures on geophysical fluid dynamics\/}.  \publ{Oxford University Press}.

\bibitem[Scinocca \& Shepherd(1992)]{scinocca1992nonlinear}
{\sc \au{Scinocca, JF} \& \au{Shepherd, TG}} \yr{1992}  \at{Nonlinear wave-activity conservation laws and {H}amiltonian structure for the two-dimensional anelastic equations}.  \jt{Journal of Atmospheric Sciences}  \bvol{49}~(1),  \pg{5--28}.

\bibitem[Shavit {\em et~al.\/}(2024)Shavit, B{\"u}hler \& Shatah]{shavit2024sign}
{\sc \au{Shavit, Michal}, \au{B{\"u}hler, Oliver} \& \au{Shatah, Jalal}} \yr{2024}  \at{Sign-indefinite invariants shape turbulent cascades}.  \jt{Physical Review Letters}  \bvol{133}~(1),  \pg{014001}.

\bibitem[Shavit {\em et~al.\/}(2025)Shavit, B{\"u}hler \& Shatah]{shavit2025turbulent}
{\sc \au{Shavit, Michal}, \au{B{\"u}hler, Oliver} \& \au{Shatah, Jalal}} \yr{2025}  \at{Turbulent spectrum of 2d internal gravity waves}.  \jt{Physical Review Letters}  \bvol{134}~(5),  \pg{054101}.

\bibitem[Veronis(1970)]{veronis1970analogy}
{\sc \au{Veronis, George}} \yr{1970}  \at{The analogy between rotating and stratified fluids}.  \jt{Annual Review of Fluid Mechanics}  \bvol{2}~(1),  \pg{37--66}.

\bibitem[Whalen {\em et~al.\/}(2020)Whalen, De~Lavergne, Naveira~Garabato, Klymak, MacKinnon \& Sheen]{whalen2020internal}
{\sc \au{Whalen, Caitlin~B}, \au{De~Lavergne, Casimir}, \au{Naveira~Garabato, Alberto~C}, \au{Klymak, Jody~M}, \au{MacKinnon, Jennifer~A} \& \au{Sheen, Katy~L}} \yr{2020}  \at{Internal wave-driven mixing: Governing processes and consequences for climate}.  \jt{Nature Reviews Earth \& Environment}  \bvol{1}~(11),  \pg{606--621}.

\bibitem[Zakharov {\em et~al.\/}(2012)Zakharov, L'vov \& Falkovich]{ZLF}
{\sc \au{Zakharov, Vladimir~E}, \au{L'vov, Victor~S} \& \au{Falkovich, Gregory}} \yr{2012} {\em Kolmogorov spectra of turbulence I: Wave turbulence\/}.  \publ{Springer Science \& Business Media}.

\end{thebibliography}

\end{document}